# Quantum Phase Transition in the Dicke Model

Moorad Alexanian

*Department of Physics and Physical Oceanography*
*University of North Carolina Wilmington, Wilmington, NC 28403-5606*

Email: alexanian@uncw.edu

**Abstract.** We consider a previously modified Jaynes-Cummings model with single-photon cavity radiation field and atomic system exchanging a squeezed photon and deduce a normal/superradiance quantum phase transition in the Dicke model of *N* atoms of arbitrary spin with independent co- and counter-rotating coupling terms.

## 1. Introduction

The Dicke model is a well-known example in quantum optics and quantum mechanics which describes a collective behavior of *N* two-level atoms interacting with a single-mode electromagnetic field in an optical cavity [1]. The Dicke Hamiltonian, a simple quantum-optical model, exhibits a zero-temperature quantum phase transition. Numerical results have been presented that at this transition the system changes from being quasi-integrable to quantum chaotic [2]. Specifically, the Dicke model with a single two-level system is called the Rabi model. The Dicke model has been generalized for spin-1 atoms, where a detailed history of the model is reviewed [3].

In this paper, we generalize a recently introduced modified Jaynes-Cummings model [4] for arbitrary spin and determine a normal/superradiance quantum phase transition in the Dicke model. This paper is arranged as follows. In Sec.2, the modified Jaynes-Cummings model is reviewed and the previous conditions on the spin states are relaxed. In Sec.3, the Dicke model is considered and the normal/superradiance, quantum phase transition is determined. Finally, Sec.4 summarizes our results.

## 2. Modified Jaynes-Cummings model

In a recent paper, we determined the quantum normal/superradiance phase transition in a modified Jaynes-Cummings model [4] with Hamiltonian

$$\hat{H}_{MJC} = \hbar\omega_c \hat{a}^\dagger \hat{a} + \hbar\omega_a \frac{\hat{\sigma}_z}{2} + \frac{\hbar\Omega}{2}(\hat{\sigma}_+ \hat{B} + \hat{\sigma}_- \hat{B}^\dagger), \qquad (1)$$

where $\omega_a = \omega_2 - \omega_1$, with $\hbar\omega_1$, $\hbar\omega_2$ are the energies of the uncoupled states $|1\rangle$ and $|2\rangle$, respectively, and $\omega_c$ is the frequency of the field mode. The creation and annihilation operators $\hat{B}$ and $\hat{B}^\dagger$, respectively, for the squeezed photons and $\hat{a}$ and $\hat{a}^\dagger$ are the photon creation and annihilation operators. The system can be in two possible states $|i\rangle$, $i = 1, 2$ with $|1\rangle$ being the ground state of the system and $|2\rangle$ being the excited state, respectively.

Hamiltonian (1) can be expressed in terms of photon creation and annihilation operators and so

$$\hat{H}_{MJC} = \hbar\omega_c \hat{a}^\dagger \hat{a} + \hbar\omega_a \frac{\hat{\sigma}_z}{2} + \frac{\hbar\Omega}{2}[\cosh(r)(\hat{\sigma}_+\hat{a} + \hat{\sigma}_-\hat{a}^\dagger) + \sinh(r)(\hat{\sigma}_-\hat{a} + \hat{\sigma}_+\hat{a}^\dagger)], \quad (2)$$

where we have chosen the squeezing operator argument $\varphi = 0$. The last term does not occur when making the rotating-wave approximation.

### A. Unitary Transformation

The Hamiltonian $\hat{H}_{MJC}$ is transformed via the unitary transformation $\hat{U}$

$$\hat{H} = \hat{U}^\dagger \hat{H}_{MJC} \hat{U}, \quad (3)$$

where

$$\hat{U} = e^{-v(\sigma_+ \hat{B}^\dagger - \sigma_- \hat{B})} \quad (4)$$

with $v$ real and $\langle \sigma_\pm \rangle = 0$ so

$$\hat{H} = \left[\hbar\omega_c(1 - v^2\hat{\sigma}_z) - \hbar v(\Omega + \omega_a v)e^{-2r}\hat{\sigma}_z\right]\hat{a}^\dagger\hat{a}$$
$$+\frac{\hbar\omega_a}{2}(v^2 + \hat{\sigma}_z) - \frac{\hbar v}{2}(\Omega + \omega_a v)e^{-2r}\hat{\sigma}_z + \hbar\omega_c v^2\cosh^2(r) - \frac{\hbar\omega_c v^2}{2}(1+\hat{\sigma}_z) \quad (5)$$
$$+\frac{\hbar v}{2}\left[\Omega\cosh(2r) - \omega_a v\sinh(2r)\right]\hat{\sigma}_z(\hat{a}^\dagger + \hat{a})^2,$$

which reduces to the previous case [4] when $\langle\hat{\sigma}_z\rangle = -1$. The Pauli matrices used to obtain (5) are

$$[\hat{\sigma}_z, \hat{\sigma}_\pm] = \pm 2\hat{\sigma}_\pm$$
$$[\hat{\sigma}_+, \hat{\sigma}_-] = \hat{\sigma}_z, \quad (6)$$

where

$$\hat{\sigma}_\pm = \frac{1}{2}(\hat{\sigma}_x \pm i\hat{\sigma}_y). \quad (7)$$

In the above derivation, we have used the leading terms in the Baker–Campbell–Hausdorff formula

$$e^{\hat{X}}\hat{Y}e^{-\hat{X}} = \hat{Y} + [\hat{X}, \hat{Y}] + \frac{1}{2!}[\hat{X}, [\hat{X}, \hat{Y}]] + \cdots \quad (8)$$

### B. Diagonalization

Consider the diagonalization of the Hamiltonian (5)

$$\hat{H} = A\hat{a}^\dagger\hat{a} + B + C(\hat{a} + \hat{a}^\dagger)^2. \quad (9)$$

This is accomplished with the aid of the Bogoliubov transformation of linear boson operators

$$\hat{a} = \cosh(\beta)\hat{b} + \sinh(\beta)\hat{b}^\dagger$$
$$\hat{a}^\dagger = \cosh(\beta)\hat{b}^\dagger + \sinh(\beta)\hat{b}. \quad (10)$$

The cancellation of the terms $\hat{b}\hat{b} + \hat{b}^\dagger\hat{b}^\dagger$ implies that

$$A\cosh(\beta)\sinh(\beta) + C\left[\cosh(\beta) + \sinh(\beta)\right]^2 = 0 \quad (11)$$

and so

$$e^{-4\beta} = \frac{A + 4C}{A}, \quad (12)$$

where

$$A = \hbar\omega_c - \hbar\omega_c v^2 \langle \hat{\sigma}_z \rangle - \hbar v e^{-2r}(\Omega + v\omega_a)\langle \hat{\sigma}_z \rangle$$
$$C = \frac{\hbar v}{2}(\Omega\cosh(2r) - v\omega_a\sinh(2r))\langle \hat{\sigma}_z \rangle \quad (13)$$

Hamiltonian (9) becomes

$$\hat{H} = B + A\sinh(\beta)\left[\sinh(\beta) - \cosh(\beta)\right] + A\left[\sinh(\beta) - \cosh(\beta)\right]^2 \hat{b}^\dagger\hat{b}, \quad (14)$$

or

$$\hat{H} = B - \frac{A}{2} + \sqrt{A(A+4C)}\,(\hat{b}^\dagger\hat{b} + 1), \quad (15)$$

where

$$B = \frac{\hbar\omega_a}{2}(v^2 + \langle\hat{\sigma}_z\rangle) - \frac{\hbar v}{2}e^{-2r}(\Omega + v\omega_a)\langle\hat{\sigma}_z\rangle + \frac{\hbar v^2\omega_c}{2}(\cosh(2r) - \langle\hat{\sigma}_z\rangle). \quad (16)$$

The constants A, B, C given by (13) and (16) reduce to the case $\langle\hat{\sigma}_z\rangle = -1$ considered in Ref.4. The quantum phase transition is characterized by $A(A + 4C) = 0$. There are two possible cases, $A = 0$ and/or $A + 4C = 0$.

### C. Quantum Phase Transition

*1. Case A=0*

One obtains for the real variable $v$ associated with the unitary transformation (4) by setting $A = 0$ in (13)

$$v = \frac{\Omega\langle\hat{\sigma}_z\rangle \pm \sqrt{\Omega^2\langle\hat{\sigma}_z\rangle^2 + 4\omega_c\langle\hat{\sigma}_z\rangle(\omega_c + \omega_a e^{-2r})e^{4r}}}{-2\langle\hat{\sigma}_z\rangle(\omega_a + \omega_c e^{2r})}, \quad (17)$$

where $\langle\hat{\sigma}_z\rangle \neq 0$. In order to remain in one quantum phase and thus avoid a bifurcation of the value of $v$, which suggests a different quantum phase for $\Omega^2 > \Omega_c^2$, we must have

$$\Omega_c^2 = \frac{-4\omega_c}{\langle\hat{\sigma}_z\rangle}(\omega_c + \omega_a e^{-2r})e^{4r}. \quad (18)$$

Therefore, the existence of a phase transition requires $\langle\hat{\sigma}_z\rangle < 0$.

*2. Case A+4C=0*

One obtains for the real variable $v$ associated with the unitary transformation (4) by setting $A+4C = 0$ and using (13)

$$v = \frac{\Omega\langle\hat{\sigma}_z\rangle \pm \sqrt{\Omega^2\langle\hat{\sigma}_z\rangle^2 + 4\omega_c\langle\hat{\sigma}_z\rangle(\omega_a + \omega_c e^{-2r})e^{-2r}}}{2\langle\hat{\sigma}_z\rangle(\omega_a + \omega_c e^{-2r})}, \tag{19}$$

As in the previous case for $A = 0$, we must have a critical point in order to remain in one quantum phase and so

$$\Omega_c^2 = -\frac{4\omega_c}{\langle\hat{\sigma}_z\rangle}(\omega_a + \omega_c e^{-2r})e^{-2r}, \tag{20}$$

where we must have $\langle\hat{\sigma}_z\rangle < 0$ and a differing quantum phase occurs for $\Omega^2 > \Omega_c^2$.

### D. Superradiance

The existence of a different quantum phase for $\Omega^2 > -4\omega_c(\omega_a + \omega_c e^{-2r})e^{-2r}/\langle\hat{\sigma}_z\rangle$ is associated with the bifurcation of the value of $v$ in (17). In order to obtain results for this new quantum state, we consider a Glauber displacement operator on the modified Jaynes-Cummings Hamiltonian (2)

$$\hat{H}_d(\alpha) = \hat{D}^\dagger(\alpha)\hat{H}_{MJC}\hat{D}(\alpha) = \hbar\omega_c(\hat{a}^\dagger + \alpha)(\hat{a} + \alpha) + \hbar\omega_a\frac{\hat{\sigma}_z}{2} + \frac{\hbar\Omega\alpha e^r}{2}(\hat{\sigma}_+ + \hat{\sigma}_-)$$
$$+ \frac{\hbar\Omega}{2}\left[\cosh(r)(\hat{\sigma}_+\hat{a} + \hat{\sigma}_-\hat{a}^\dagger) + \sinh(r)(\hat{\sigma}_-\hat{a} + \hat{\sigma}_+\hat{a}^\dagger)\right] \tag{21}$$

where

$$\hat{D}(\alpha) = e^{\alpha(\hat{a}^\dagger - \hat{a})} \tag{22}$$

and $\alpha$ real. The eigenstates of the atomic part of the Hamiltonian (21), viz., $\hbar\omega_a\hat{\sigma}_z/2 + \hbar\alpha e^r\Omega\hat{\sigma}_x/2$, are

$$|\tilde{\uparrow}\rangle = \cos(\theta)|\uparrow\rangle + \sin(\theta)|\downarrow\rangle, \qquad |\tilde{\downarrow}\rangle = -\sin(\theta)|\uparrow\rangle + \cos(\theta)|\downarrow\rangle \tag{23}$$

with $\tan(2\theta) = \alpha\Omega e^r/\omega_a$. The new transition frequency is given by $\tilde{\Omega} = \sqrt{\omega_a^2 + \alpha^2\Omega^2 e^{2r}}$, where $\pm\hbar\tilde{\Omega}/2$ are the eigenvalues of the atomic part of the Hamiltonian (21). The relation between Pauli matrices is as follows

$$\hat{\sigma}_x = \cos(2\theta)\hat{\tau}_x + \sin(2\theta)\hat{\tau}_z$$
$$\hat{\sigma}_y = \hat{\tau}_y \tag{24}$$
$$\hat{\sigma}_z = -\sin(2\theta)\hat{\tau}_x + \cos(2\theta)\hat{\tau}_z.$$

In terms of Pauli matrices $\hat{\tau}_{x,y,z}$ in the $|\tilde{\uparrow}\rangle, |\tilde{\downarrow}\rangle$ basis, (21) becomes

$$\tilde{H}_d(\alpha) = \hbar\omega_c\hat{a}^\dagger\hat{a} + \hbar\omega_c\alpha^2 + \frac{\hbar\Omega^2 e^{2r}}{8\omega_c}\hat{\tau}_z + \frac{\hbar\Omega\cos(2\theta)e^r}{4}(\hat{a} + \hat{a}^\dagger)\hat{\tau}_x. \tag{25}$$

The Hamiltonian (25) is of the generic Rabi form

$$\hat{H}_g = J\hat{a}^\dagger\hat{a} + K + L\hat{\tau}_z + M(\hat{a} + \hat{a}^\dagger)\hat{\tau}_x, \tag{26}$$

where *J, K, L, M* are real constants, that can be diagonalized with the aid of the unitary transformation $\hat{S}(\mu) = \exp[\mu(\hat{a} + \hat{a}^\dagger)(\hat{\sigma}_+ - \hat{\sigma}_-)]$ with real $\mu$ and so

$$\hat{S}^\dagger(\mu)\hat{H}_g\hat{S}(\mu) = J\hat{a}^\dagger\hat{a} + K + L\hat{\tau}_z - 2\mu(M + \mu L)(\hat{a} + \hat{a}^\dagger)^2\hat{\tau}_z. \quad (27)$$

Result (27) follows by keeping the lowest order terms in the Baker–Campbell–Hausdorff formula (8). The excitation energy follows from (15) and so

$$\epsilon = J\sqrt{1 + \cos^2(2\theta)\langle\hat{\tau}_z\rangle}. \quad (28)$$

Now

$$\mu = \frac{\tilde{\lambda}}{\tilde{\Omega}} = \frac{-M}{2L} = -\frac{\omega_c \cos(2\theta)}{\Omega e^r}, \quad (29)$$

where $-\tilde{\lambda}$ is the coefficient of the $(\hat{a} + \hat{a}^\dagger)\hat{\tau}_x$ term in (25) and $\tilde{\Omega}/2$ is the coefficient of the $\hat{\tau}_z$ term in (25) [5] and

$$\cos(2\theta) = \frac{\omega_a}{\sqrt{\omega_a^2 + \alpha^2\Omega^2 e^{2r}}}. \quad (30)$$

## 3. Dicke model

The Dicke model Hamiltonian is given by

$$\hat{H}_D = \hbar\omega_c\hat{a}^\dagger\hat{a} + \hbar\omega_a\frac{\hat{S}_z}{2} + \frac{\hbar\lambda_-}{\sqrt{2N}}(\hat{a}\hat{S}_+ + \hat{a}^\dagger\hat{S}_-) + \frac{\hbar\lambda_+}{\sqrt{2N}}(\hat{a}\hat{S}_- + \hat{a}^\dagger\hat{S}_+), \quad (31)$$

where $\omega_c$ and $\omega_a$ are cavity and atomic frequencies. The interaction terms with $\lambda_\pm$ (given by the Raman transition rate) are the coupling constants for the co- and counter-rotating terms [3]. The operator $\hat{a}$ is the cavity mode annihilation operator, where $\hat{S}_z$ and $\hat{S}_\pm$ are collective atomic spin given by

$$\hat{S}_{z,\pm} = \sum_{k=1}^{N} \hat{S}_{z,\pm}^k, \quad (32)$$

where $\hat{S}_{z,\pm}^k$ is the spin operator for the *k*-atom and *N* is the number of atoms. The collective atomic spin operators satisfy

$$[\hat{S}_z, \hat{S}_\pm] = \pm 2\hat{S}_\pm$$
$$[\hat{S}_+, \hat{S}_-] = \hat{S}_z, \quad (33)$$

just as the Pauli matrices in (6). Accordingly, results in Sec.1 apply directly to the Dicke model with the replacements

$$(\Omega/2)\cosh(r) \to \lambda_-/\sqrt{2N}$$
$$(\Omega/2)\sinh(r) \to \lambda_+/\sqrt{2N} \quad (34)$$
$$\hat{\sigma}_{z,\pm} \to \hat{S}_{z,\pm}.$$

Therefore, and so

$$0 \leq \frac{\lambda_+}{\lambda_-} \leq 1, \qquad (36)$$

with $\lambda_- \geq \lambda_+ \geq 0$ or $\lambda_- \leq \lambda_+ \leq 0$.

$$e^{2r} = \frac{\lambda_- + \lambda_+}{\lambda_- - \lambda_+}$$
$$\Omega^2 = \frac{2}{N}(\lambda_-^2 - \lambda_+^2), \qquad (35)$$

### A. Quantum phase transition

The quantum phase transition is characterized by $A(A + 4C) = 0$.

*1. Case A=0*

Condition (18) gives

$$\frac{2}{N}\lambda_-^2(1 - \lambda_+/\lambda_-)^2 = \frac{-4\omega_c}{\langle \hat{S}_z \rangle}\left(\omega_c + \omega_a\frac{1 - \lambda_+/\lambda_-}{1 + \lambda_+/\lambda_-}\right)\left(\frac{1 + \lambda_+/\lambda_-}{1 - \lambda_+/\lambda_-}\right)^2, \qquad (37)$$

where $\lambda_-$ is determined for a given value of $\lambda_+/\lambda_-$. Note there is no phase transition for the case $\lambda_- = \lambda_+$ as noted for the Rabi model [4]. For the Jaynes-Cummings model, $\lambda_+ = 0$, $N = 1$, and $\langle \hat{S}_z \rangle = -1$, we have [4]

$$2\lambda_-^2 = 4\omega_c(\omega_a + \omega_c). \qquad (38)$$

However, there is always a phase transition provided $\lambda_- \neq \lambda_+$. In the thermodynamic limit $N \to \infty$, there is no phase transition for $-\infty < \langle \hat{S}_z \rangle < 0$ since $0 \leq \lambda_+/\lambda_- \leq 1$.

*2. Case A+4C=0*

Condition (20) gives

$$\frac{2}{N}\lambda_-^2(1 + \lambda_+/\lambda_-)^2 = \frac{-4\omega_c}{\langle \hat{S}_z \rangle}\left(\omega_a + \omega_c\frac{1 - \lambda_+/\lambda_-}{1 + \lambda_+/\lambda_-}\right), \qquad (39)$$

where $\lambda_-$ is determined for a given value of $\lambda_+/\lambda_-$. The Jaynes-Cummings critical point [4] is given by $\lambda_+ = 0$, $N = 1$, $\langle \hat{S}_z \rangle = -1$, and so (39) becomes

$$2\lambda_-^2 = 4\omega_c(\omega_a + \omega_c). \qquad (40)$$

The Rabi critical point [4] is given by $\lambda_- = \lambda_+$, $N = 1$, $\langle \hat{S}_z \rangle = -1$, and so (39) becomes

$$\lambda_-^2 = \frac{1}{2}\omega_c\omega_a. \qquad (41)$$

In particular, in the thermodynamic limit $N \to \infty$, there is no phase transition for $-\infty < \langle \hat{S}_z \rangle < 0$ since $0 \leq \lambda_+/\lambda_- \leq 1$.

### B. Ground-state energy

The ground-state energy is given by

$$\epsilon = \sqrt{A(A+4C)}, \quad (42)$$

$$A = \hbar\omega_c - \frac{\hbar\lambda_-^2(1-\lambda_+^2/\lambda_-^2)}{2\left[\omega_a + \omega_c \frac{1-\lambda_+/\lambda_-}{1+\lambda_+/\lambda_-}\right]^2}\left\{\omega_c\left[1+2\left(\frac{1-\lambda_+/\lambda_-}{1+\lambda_+/\lambda_-}\right)^2\right]+3\omega_a\left(\frac{1-\lambda_+/\lambda_-}{1+\lambda_+/\lambda_-}\right)\right\}\frac{\langle S_z\rangle}{N} \quad (43)$$

$$A + 4C = \hbar\omega_c + \frac{\hbar\lambda_-^2(1+\lambda_+/\lambda_-)^2}{2\left(\omega_a + \omega_c\frac{1-\lambda_+/\lambda_-}{1+\lambda_+/\lambda_-}\right)}\frac{\langle S_z\rangle}{N},$$

where

where we recover the results for the Rabi model ($\lambda_+ = \lambda_-$) and the Jaynes-Cummings model ($\lambda_+ = 0$), for $\langle \hat{S}_z \rangle = -1$ and $N = 1$ [4].

### C. Superradiance

The existence of a different quantum phase for $\Omega^2 > 4\omega_c(\omega_a + \omega_c e^{-2r})e^{-2r}$ is associated with the bifurcation of the value of $v$ in (19). In order to obtain results for this new quantum state, we consider a Glauber displacement operator on the Dicke model Hamiltonian (31). The procedure is just as in Section 2 for the modified Jaynes-Cummings model resulting in an excitation energy analogous to (28) but with the corresponding replacement (34) and with excitation energy

$$\epsilon = \hbar\omega_c\sqrt{1 + \cos^2(2\theta)\langle \hat{S}_z\rangle}, \quad (44)$$

where

$$\cos^2(2\theta) = \frac{\omega_a^2}{\omega_a^2 + 2\alpha^2\lambda_-^2(1+\lambda_+/\lambda_-)^2/N} \quad (45)$$

We define $\alpha^2/N$ via

$$\cos^2(2\theta) = \frac{4\omega_c^2}{\lambda_-^4(1+\lambda_+/\lambda_-)^4 N}\left[\omega_a + \omega_c\left(\frac{1-\lambda_+/\lambda_-}{1+\lambda_+/\lambda_-}\right)\right]^2 \quad (46)$$

The excitation energy (44) reduces to the Rabi case ($\lambda_+ = \lambda_-$), the Jaynes-Cummings case ($\lambda_+ = 0$) for $N = 1$ and $\langle \hat{S}_z \rangle = -1$. The excitation energy (44) exists in the thermodynamic limit $N \to \infty$ with the ratio $\langle \hat{S}_z \rangle/N$ is negative and finite. Note that there is no phase transition in the thermodynamic limit $N \to \infty$ with $\langle \hat{S}_z \rangle$ finite.

## 4. Conclusions

A modified Jaynes-Cummings model is generalized to the Dicke model of $N$ atoms of arbitrary spin with independent co- and counter-rotating coupling terms. The model gives rise to a normal/superradiance quantum phase transition as in the Jaynes-Cummings and Rabi models as shown in a previous publication.